\documentclass[pra,twocolumn,showpacs]{revtex4}
\usepackage{graphicx}
\usepackage{amsmath}
\usepackage{amssymb}
\newcommand{\sss}{\scriptstyle}
\def\lsim{\
  \lower-1.2pt\vbox{\hbox{\rlap{$<$}\lower5pt\vbox{\hbox{$\sim$}}}}\ }
\def\gsim{\
  \lower-1.2pt\vbox{\hbox{\rlap{$>$}\lower5pt\vbox{\hbox{$\sim$}}}}\ }

\begin{document}

\title{Point bosons in a one-dimensional box: \\ the ground state,  excitations and  thermodynamics}

\author{Maksim Tomchenko}
\email{mtomchenko@bitp.kiev.ua}
\affiliation{Bogolyubov Institute for Theoretical Physics,
        14-b Metrolohichna Str., Kyiv 03680, Ukraine}

\date{\today}
\begin{abstract}
We determine the ground-state energy and the effective dispersion law
for a one-dimensional system of point bosons under zero boundary conditions.
The ground-state energy is close to the value
for a periodic system.
But the dispersion law is essentially different from that for a periodic system, if the coupling
is weak (weak interaction or high concentration) or intermediate.
We propose also a new method for construction of the thermodynamics for
a gas of point bosons. It turns out that the difference
in the dispersion laws of systems with periodic and
zero boundary  conditions does not lead to a difference in the thermodynamic quantities.
In addition, under zero boundary conditions, the microscopic  sound velocity does not coincide
with the macroscopic one. This means that either the method of determination of $k$
in the dispersion law $E(k)$ is unsuitable
or the low-energy excitations are not phonons.
       \end{abstract}
\pacs{67.25.dt,  67.25.dr}

\maketitle

 \section{Introduction}
Systems of many particles with contact interaction were studied in a lot of works, starting with the work by
Bethe \cite{bethe} (see  other references  in monographs \cite{gaudinm,baxter1989,sutherland2004}
and recent reviews \cite{batchelor2014,jiang2015}).
Models for spinless bosons with point interaction were constructed in several main works: these are
the Girardeau model \cite{girardeau1960} for impenetrable particles under periodic boundary conditions (BCs),
the Lieb--Liniger model \cite{ll1963} and the Lieb model \cite{lieb1963} for penetrable particles
under periodic BCs, and  Gaudin solutions \cite{gaudin1971}
for zero BCs. Moreover, some results for the ground state under zero BCs
were obtained by Batchelor et al.  \cite{batchelor2005}.
The influence of the boundaries on the ground state of a system of point
fermions was studied as well \cite{woynarovich1985,batchelor2006}.

In the present work, we investigate the ground-state, the dispersion law, and
the  thermodynamics for spinless point bosons under zero BCs.
The two following results are basic.
We give a new method for construction of the thermodynamics for point bosons
and find out that the dispersion law under zero BCs
differs from that under periodic BCs. However, under zero BCs,
the low-energy elementary excitations are not phonons,
if we determine $k$ as in the present work. Therefore, the curve $E(k)$
under zero BCs is the ``dispersion law'' in the effective sense.
The nonphononicity arouses the questions, which are considered in Section VII.
The dispersion relations for  $SU(2)$ point bosons under periodic and zero BCs
were found in \cite{ying2001,ying2003}. But the results for periodic and zero BCs
are given at different system parameters; therefore, it is hard to see whether
dispersion relations depend on boundaries at large $N, L$.

 \section{Initial equations}
We now recall briefly the basic equations. They are presented in the elegant and exact form
in the works by Gaudin \cite{gaudin1971,gaudinm}.

Let us consider $N$ bosons, which occupy a one-dimensional (1D)
interval of length $L$  and interact by means of a binary
repulsive potential in the form of a delta-function. The interval
can be closed or nonclosed. The Schr\"{o}dinger equation for such
system is usually written in the form \cite{ll1963}
\begin{equation}
 -\sum\limits_{j}\frac{\partial^{2}}{\partial x_{j}^2}\Psi + 2c\sum\limits_{i<j}
\delta(x_{i}-x_{j})\Psi=E\Psi.
     \label{1} \end{equation}
Under periodic BCs, the solution of this equation for the  domain
$x_{1}\leq x_{2}\leq\ldots \leq x_{N}$ is the Bethe ansatz \cite{ll1963,gaudin1971}
\begin{equation}
 \Psi_{\{k \}}(x_{1},\ldots,x_{N})=\sum\limits_{P}a(P)e^{i\sum\limits_{l=1}^{N} k_{P_{l}}x_{l}},
      \label{2} \end{equation}
where  $k_{P_{l}}$ is equal to one of $k_{1},\ldots,k_{N}$, and $P$ means all permutations of $k_{l}$. The coefficients
$a(P)$ were determined in \cite{lieb1963}. Under zero BCs, the solution is a superposition of plane waves
(\ref{2}) and all possible reflected waves \cite{gaudin1971}.

The energy of a system of point bosons
\begin{eqnarray}
 E=k_{1}^{2}+k_{2}^{2}+\ldots +k_{N}^{2}.
     \label{3} \end{eqnarray}
Since the energy is completely determined by the values of $k_{l}$, it is sufficient
to find the corresponding $k_{l}$ in order to obtain the ground-state energy $E_{0}$ and the dispersion law.

For a periodic system, the equations for $k_{l}$ take the form \cite{yangs1969,gaudin1971}
\begin{eqnarray}
Lk_{i}=2\pi I^{p}_{i}-2\sum\limits_{j=1}^{N}\arctan{\frac{k_{i}-k_{j}}{c}},
     \label{4} \end{eqnarray}
\begin{equation}
I^{p}_{i}=n_{i}+i-\frac{N+1}{2}, \quad i=1,\ldots, N,
     \label{5} \end{equation}
where $n_{i}$ are integers.   For the ground state, $n_{i}=0$ for all $i$.
For $I^{p}_{1}<I^{p}_{2}<\ldots <I^{p}_{N},$ system (\ref{4}) has the unique real
solution  $\{k_{i}\}$ \cite{yangs1969}.
Using the equality
\begin{equation}
\arctan{\alpha}=(\pi/2)sgn(\alpha)-\arctan{(1/\alpha)},
     \label{6} \end{equation}
Eqs. (\ref{4}) can be rewritten in the equivalent form \cite{gaudin1971}
\begin{eqnarray}
Lk_{i}= 2\pi n_{i}+2\sum\limits_{j=1}^{N}\arctan{\frac{c}{k_{i}-k_{j}}}|_{j\neq i}.
     \label{7} \end{eqnarray}
The solutions of (\ref{4}) and (\ref{7}) are the  collections of $\{k_{i}\}$, for which
$k_{j}\neq k_{i}$ for any $j\neq i$.
Below, we assume the ordering $k_{1}<k_{2}<\ldots < k_{N}$, at which the equality
$k_{j}=-k_{N+1-j}$ holds for the ground state.

For a system with zero BCs, the equations for $k_{l}$ take the form \cite{gaudin1971,gaudinm}
\begin{eqnarray}
L|k_{i}|&=&\pi n_{i}+\sum\limits_{j=1}^{N}\left (\arctan{\frac{c}{|k_{i}|-|k_{j}|}}\right.
\nonumber\\  &+& \left.
\arctan{\frac{c}{|k_{i}|+|k_{j}|}}\right )|_{j\neq i}, \quad i=1,\ldots, N,
     \label{8} \end{eqnarray}
where $n_{i}$ are integers, and $n_{i}\geq 1$.
With regard for (\ref{6}), these equations can be written in the form
\begin{eqnarray}
L|k_{i}| &=&\pi I^{z}_{i}-\sum\limits_{j=1}^{N}\left (\arctan{\frac{|k_{i}|-|k_{j}|}{c}}\right. \nonumber\\  &+& \left.
\arctan{\frac{|k_{i}|+|k_{j}|}{c}}\right )\left. \right |_{j\neq i}, \quad i=1,\ldots, N,
     \label{9} \end{eqnarray}
\begin{equation}
I^{z}_{i}=n_{i}+i-1.
     \label{10} \end{equation}
The quantities $k_{l}$ are commonly ordered in the following way: $0< |k_{1}|< |k_{2}|< \ldots < |k_{N}|$.
For brevity, we will write $k_{l}$ instead of $|k_{l}|$. In what follows,
$k_{l}$ means $|k_{l}|$ everywhere under zero BCs. We denote $n=N/L$ and $\gamma =c/n$.

  \section{Ground state of bosons in a box}
For point bosons in a box, the ground state is described
by Eqs. (\ref{8}) -- (\ref{10}) with $n_{i}=1$.
We can verify that $n_{i}=1$ in the following way (the strict proof is absent, as far as we know).
Since $k_{1}>0$ is the smallest quasimomentum, the equation
\[Lk_{1}=\pi n_{1}+\sum\limits_{j=2}^{N}\left (\arctan{\frac{c}{k_{1}-k_{j}}} +
\arctan{\frac{c}{k_{1}+k_{j}}}\right )\]
is satisfied only for $n_{1}\geq 1$. Otherwise, the right-hand side is  negative,
but the left-hand side must be positive. Therefore, the minimum value of $n_{1}$ is equal to $1$.
The symmetry-based reasoning indicates that the ordering $k_{1}< k_{2}< \ldots < k_{N}$
assumes $ n_{1}\leq n_{2}\leq \ldots \leq n_{N}$. Therefore, $n_{i}\geq 1$ for all $i$.
The smallest $n_{i}$ are $n_{i}= 1$. In addition, in the limit $c\rightarrow 0$
we should obtain the momenta of free particles $k_{i}=\pi/L$,
which also requires $n_{i}= 1$ \cite{gaudin1971}.
We have studied Eq. (\ref{8}) numerically by the Newton method and have found
no states with the energy, which is lower then the energy of the state with $n_{i}= 1$ for any $i$.
We will assume that, for any $c,$ the ground state corresponds to $n_{i}= 1$ for all $i$.

1) Ultraweak coupling ($c\rightarrow 0$). In this case,
$|c/(k_{i}-k_{j})|\ll 1$ for all $i\neq j$. Therefore, relation (\ref{8}) for the ground state
changes into
\begin{eqnarray}
Lk_{i}=\pi +\sum\limits_{j=1}^{N}\left (\frac{c}{k_{i}-k_{j}} +
\frac{c}{k_{i}+k_{j}}\right )|_{j\neq i}.
     \label{12} \end{eqnarray}
Making use of the change $k_{i}=\sqrt{c/L}\cdot (q_{i}+\pi/\sqrt{c L}),$ we obtain the equation
\begin{eqnarray}
q_{i}=\sum\limits_{j=1}^{N}\left (\frac{1}{q_{i}-q_{j}} +
\frac{1}{q_{i}+q_{j}+2\pi/\sqrt{c L}}\right )|_{j\neq i}.
     \label{13} \end{eqnarray}
As $c\rightarrow 0$, we obtain
\begin{eqnarray}
q_{i}=\sum\limits_{j=1}^{N}\frac{1}{q_{i}-q_{j}}|_{j\neq i} + \frac{(N-1)\sqrt{c L}}{2\pi}+
O(c).
     \label{14} \end{eqnarray}
We set $q_{i}=q^{(N)}_{i}+(N-1)\sqrt{c L}/(2\pi)$, then
\begin{equation}
q^{(N)}_{i}=  \sum\limits_{j=1}^{N}\frac{1}{q^{(N)}_{i}-q^{(N)}_{j}} |_{j\neq i}.
     \label{15} \end{equation}
As was noticed by Gaudin \cite{gaudin1971}, it is the equation for roots of the Hermite
polynomial $H_{N}(q)$. This can be verified
by substituting the polynomial
\begin{equation}
H_{N}(q)= 2^{N}\prod\limits_{j=1}^{N}(q-q^{(N)}_{j})
     \label{16} \end{equation}
into the equation for the Hermite polynomials
\begin{equation}
\ddot{H}_{N}(q)-2q\dot{H}_{N}(q)+2NH_{N}(q)= 0
     \label{17} \end{equation}
and setting $q=q^{(N)}_{i}$.
The roots $q^{(N)}_{i}$ of Hermite polynomials satisfy the relations
\begin{equation}
\sum\limits_{j=1}^{N}q^{(N)}_{i}=0,
     \label{19} \end{equation}
\begin{equation}
\sum\limits_{j=1}^{N}\left (q^{(N)}_{i} \right )^{2}|_{N\gg 1}=N(N-1)/2 +O(N^{-1})
     \label{20} \end{equation}
(to obtain the last property, we found $q^{(N)}_{i}$ numerically for $N\leq 100$;
values of $q^{(N)}_{i}$ for $N\leq 20$ are given in \cite{abramovitz1972}).
Using these formulae, we find the ground-state energy for $N\gg 1$:
\begin{eqnarray}
E_{0}=\sum\limits_{i=1}^{N}k_{i}^{2}= \frac{N\pi^{2}}{L^{2}}+\frac{3(N-1)c n}{2}+\xi,
     \label{18} \end{eqnarray}
where $\xi$ is small.
Formula (\ref{18}) was previously obtained in \cite{batchelor2005}.
This formula can be obtained also
by algebraic transformations. In this way, we get
\begin{eqnarray}
\xi =  \frac{\pi c}{L^{2}}\sum\limits_{i,j=1}^{N}\frac{2k_{i}}{k_{i}-k_{j}}\left (
\frac{1}{k_{i}+k_{j}}-\frac{L}{2\pi}  \right )|_{j\neq i}.
     \label{18b} \end{eqnarray}
The numerical solution of Eqs. (\ref{8}) by the Newton method shows that
the correction $\xi$  in (\ref{18}) can be neglected if $\gamma \ll 1/N$.
For $\gamma \gsim 1/N,$ the quantity $E_{0}$ is close to the Bogolyubov solution
for a periodic system: $E_{0}\approx Ncn$.

2) Weak and intermediate couplings, $N^{-1}\lsim \gamma \lsim 10$.
In \cite{gaudin1971}, it was asserted that $E_{0}$ for a  $(N,L,c)$-system with boundaries
coincides with a half of the energy of the periodic $(2N,2L,c)$-system with $n_{i}=-1$ for $i=1,\ldots,N$
and $n_{i}=1$ for $i=N+1,\ldots,2N$. This is not quite so, because the first system
is described by Eq. (\ref{8}), whereas, for the positive quasimomenta of the second system, relation (\ref{7})
yields, after the appropriate reenumeration, the equation
\begin{eqnarray}
Lk_{i}&=& \sum\limits_{j=1}^{N}\left (\arctan{\frac{c}{k_{i}-k_{j}}} +
\arctan{\frac{c}{k_{i}+k_{j}}}\right )|_{j\neq i}\nonumber\\  &+&
 \pi +\arctan{\frac{c}{2k_{i}}}, \quad i=1,\ldots, N,
     \label{21} \end{eqnarray}
which differs from (\ref{8}) by the term $\arctan{(c/2k_{i})}$. Therefore, there is no exact correspondence
between the $(N,L,c)$-system with boundaries and the periodic $(2N,2L,c)$-system.

The value of $E_{0}$ was already calculated in \cite{gaudin1971,gaudinm,batchelor2005}.
But M. Gaudin and M. Batchelor et al. used a method assuming the proximity of the values of $E_{0}$
under periodic and zero BCs. However, these $E_{0}$ may strongly differ from each other.
Therefore, we use a method similar to that in \cite{ll1963},
which does not require the proximity of the solution for $E_{0}$ to that under  periodic BCs.

Under zero BCs, the quasimomenta $k_{i}$ satisfy Eq. (\ref{8}).
For $N\gg 1,$ $k_{i}$ vary smoothly as $i$ increases. Therefore,
it is convenient to pass in (\ref{8}) and (\ref{9}) from summation to integration.
Let us consider  $k_{i}$ as a function of  $\varsigma =  i/N$:
$k_{i}=\upsilon(i/N)= \upsilon(\varsigma)$. Then
\begin{equation}
k_{i+1}-k_{i}=N^{-1}(\partial \upsilon/ \partial \varsigma)|_{\varsigma=i/N}\equiv 1/\rho(k_{i}).
     \label{22} \end{equation}
Since $dk=d\upsilon=(d\upsilon(\varsigma)/d\varsigma)d\varsigma=(N/\rho(k))d\varsigma$, we have
$d\varsigma =\rho(k)dk/N$. This yields
\begin{equation}
\int\limits_{k_{1}}^{k_{N}}\rho(k)dk=N\int\limits_{0}^{1}d\varsigma=N,
     \label{23} \end{equation}
\begin{equation}
\sum\limits_{i=1}^{N}f(k_{i})=\frac{1}{{\sss \triangle} \varsigma}
\int\limits_{0}^{1}f(\upsilon(\varsigma))d\varsigma=\int\limits_{k_{1}}^{k_{N}}f(k)\rho(k)dk,
     \label{24} \end{equation}
where $f(k)$ is any function, and ${\sss \triangle } \varsigma =  1/N$.
This gives the normalization for the density of states $\rho(k)$ and
the simple rule for the transition from the summation over
$k_{i}$ to the integration. If we set $f(k)=1$ in (\ref{24}), we obtain (\ref{23}).

With regard to (\ref{24}), Eq. (\ref{9}) can be written in the form
\begin{equation}
Lk_{i}=\pi i-\int\limits_{k_{1}}^{k_{N}}\rho(k)dk\left (\arctan{\frac{k_{i}-k}{c}}+
\arctan{\frac{k_{i}+k}{c}} \right )|_{k\neq k_{i}},
     \label{25} \end{equation}
where  $\ i=1,\ldots, N$, or
\begin{align}
& Lk_{i}-\arctan{\frac{2k_{i}}{c}}=\pi i \nonumber\\
& - \int\limits_{k_{1}}^{k_{N}}\rho(k)dk\left (\arctan{\frac{k_{i}-k}{c}}+
\arctan{\frac{k_{i}+k}{c}} \right ).
     \label{26} \end{align}
Considering $|k_{i+1}-k_{i}|$ small, we subtract the $i$th equation in (\ref{26}) from the $(i+1)$th one.
In view of (\ref{22}), we obtain $N-1$ equations for $\rho(k_{i})$. They can be written as the integral
equation
\begin{eqnarray}
&&\pi\rho(q)-\int\limits_{k_{1}}^{k_{N}}\rho(k)dk\left (\frac{c}{c^{2}+(k-q)^{2}}+\frac{c}{c^{2}+(k+q)^{2}}
\right ) \nonumber\\ && =L-\frac{2c}{c^{2}+4q^{2}},
     \label{27} \end{eqnarray}
where $q\in [k_{1},k_{N}]$.  Since $k_{1}$ is unknown and can be separated from $0$ by a gap,
we write additionally the equation for $k_{1},$
\begin{equation}
Lk_{1}=\pi -\int\limits_{k_{2}}^{k_{N}}\rho(k)dk\left (\arctan{\frac{k_{1}-k}{c}}+
\arctan{\frac{k_{1}+k}{c}} \right ),
     \label{28} \end{equation}
where $k_{2}=k_{1}+1/\rho(k_{1})$. Equations (\ref{23}), (\ref{27}),
and (\ref{28}) set the complete system of equations for $\rho(k)$, $k_{1},$ and $k_{N}$.

Let us compare them with the equations for the ground state of a periodic system.
Let $N$ be even.
For the \textit{positive} quasimomenta $\tilde{k}_{1}<\tilde{k}_{2}< \ldots < \tilde{k}_{N/2}$
 of a periodic system, the equations read
\begin{equation}
\int\limits_{\tilde{k}_{1}}^{\tilde{k}_{N/2}}\rho_{p}(\tilde{k})d\tilde{k}=N/2,
     \label{30} \end{equation}
\begin{align}
 L\tilde{k}_{i}=&-\int\limits_{\tilde{k}_{1}}^{\tilde{k}_{N/2}}2\rho_{p}(\tilde{k})d\tilde{k}
\left (\arctan{\frac{\tilde{k}_{i}-\tilde{k}}{c}}+\arctan{\frac{\tilde{k}_{i}+\tilde{k}}{c}} \right )
 \nonumber\\ & +\pi (2i-1), \quad i=1,\ldots, N/2.
     \label{29} \end{align}
Equation (\ref{29}) follows from (\ref{4}) and (\ref{5})
and yields
\begin{eqnarray}
 2\pi\rho_{p}(\tilde{q})&-&\int\limits_{\tilde{k}_{1}}^{\tilde{k}_{N/2}}2\rho_{p}(\tilde{k})d\tilde{k}
\left (\frac{c}{c^{2}+(\tilde{k}-\tilde{q})^{2}}\right.
 \nonumber\\  &+&\left. \frac{c}{c^{2}+(\tilde{k}+\tilde{q})^{2}}
\right )=L,
     \label{31} \end{eqnarray}
\begin{align}
L\tilde{k}_{1}=&-\int\limits_{\tilde{k}_{2}}^{\tilde{k}_{N/2}}2\rho_{p}(\tilde{k})d\tilde{k}
\left (\arctan{\frac{\tilde{k}_{1}-\tilde{k}}{c}}+\arctan{\frac{\tilde{k}_{1}+\tilde{k}}{c}} \right )
  \nonumber\\ & + \pi-2\arctan{\frac{2\tilde{k}_{1}}{c}},
     \label{32} \end{align}
$\tilde{k}_{2}=\tilde{k}_{1}+1/\rho_{p}(\tilde{k}_{1})$.
Equations (\ref{30}), (\ref{31}), and (\ref{32}) form the
complete system of equations for a periodic system,
which is written in the form of the equations for a system with boundaries.
We now make changes
$2\rho_{p}(\tilde{k})=\tilde{\rho}(\tilde{k})$, $\tilde{k}_{N/2}=\bar{k}_{N}$. Then the equations take the form
\begin{equation}
\int\limits_{\tilde{k}_{1}}^{\bar{k}_{N}}\tilde{\rho}(\tilde{k})d\tilde{k}=N,
     \label{33} \end{equation}
\begin{eqnarray}
\pi\tilde{\rho}(\tilde{q})-\int\limits_{\tilde{k}_{1}}^{\bar{k}_{N}}d\tilde{k}
\left (\frac{\tilde{\rho}(\tilde{k})c}{c^{2}+(\tilde{k}-\tilde{q})^{2}}+
\frac{\tilde{\rho}(\tilde{k})c}{c^{2}+(\tilde{k}+\tilde{q})^{2}}
\right )=L,
     \label{34} \end{eqnarray}
\begin{align}
L\tilde{k}_{1}=&-\int\limits_{\tilde{k}_{2}}^{\bar{k}_{N}}\tilde{\rho}(\tilde{k})d\tilde{k}
\left (\arctan{\frac{\tilde{k}_{1}-\tilde{k}}{c}}+\arctan{\frac{\tilde{k}_{1}+\tilde{k}}{c}} \right )
 \nonumber\\ &  +\pi-2\arctan{\frac{2\tilde{k}_{1}}{c}},
     \label{35a} \end{align}
$\tilde{k}_{2}=\tilde{k}_{1}+2/\tilde{\rho}(\tilde{k}_{1})$. Equation (\ref{35a}) can be written as
\begin{eqnarray}
L\tilde{k}_{1}&=&\pi+A_{1}-\int\limits_{\tilde{k}_{1}+1/\tilde{\rho}(\tilde{k}_{1})}^{\bar{k}_{N}}\tilde{\rho}(\tilde{k})d\tilde{k}
\left (\arctan{\frac{\tilde{k}_{1}-\tilde{k}}{c}}\right. \nonumber\\
&+&\left. \arctan{\frac{\tilde{k}_{1}+\tilde{k}}{c}} \right ),
     \label{35} \end{eqnarray}
\begin{align}
A_{1}\approx & \arctan{\left (\frac{2\tilde{k}_{1}}{c}+\frac{1.5}{c\tilde{\rho}(\tilde{k}_{1})}\right )}
\nonumber\\ -&
2\arctan{\frac{2\tilde{k}_{1}}{c}}-\arctan{\frac{1.5}{c\tilde{\rho}(\tilde{k}_{1})}}.
     \label{35b} \end{align}
Equations (\ref{33}), (\ref{34}),  (\ref{35}) differ from
(\ref{23}), (\ref{27}), (\ref{28}) only by two terms.
Equations (\ref{27}) and Eq. (\ref{35}) include the terms
$-\frac{2c}{c^{2}+4q^{2}}$ and $A_{1}$, which are absent,
respectively, in (\ref{34}) and (\ref{28}). The term
$-\frac{2c}{c^{2}+4q^{2}}$ enters the combination
$L-\frac{2c}{c^{2}+4q^{2}}$. As $N$ increases, the value of $L$
increases as well (at a fixed density $n$). Therefore,
 the quantity $-\frac{2c}{c^{2}+4q^{2}}$
gives an arbitrarily small contribution in the limit $N\rightarrow \infty.$  The quantity $A_{1}$
enters the combination $L\tilde{k}_{1}-\pi-A_{1}$.
The numerical analysis shows that, in the regime  $\gamma \gg 1/N^{2}$ and $N\geq 1000,$ the relations
$c\tilde{\rho}(\tilde{k}_{1}) \gg 1$ and $\tilde{k}_{1}/c \ll 1$ hold, from whence $|A_{1}|\ll \pi$
($\tilde{k}_{1}$ can be estimated as $\tilde{k}_{1}\sim \tilde{k}_{N/2}/N$).
In the regime $\gamma \gg 1/N,$ the relation $|A_{1}|\ll L\tilde{k}_{1}$ is valid as well.
Therefore, the quantity $A_{1}$ can be neglected.

This means that, for $\gamma \gg 1/N,$ the distinction between systems
(\ref{33}), (\ref{34}), (\ref{35}) and (\ref{23}), (\ref{27}), (\ref{28})
is negligible. Therefore, $k_{1}=\tilde{k}_{1}$, $k_{N}=\tilde{k}_{N/2},$
and $\rho(k)= \tilde{\rho}(k)=2\rho_{p}(k)$. Thus,
for a system with zero BCs, $\rho(k)$ is 2 times larger and $k_{N}$ is the same, as compared with $\rho(k)$ and $k_{N}$
for a system with periodic BCs and the same $c, N,$ $L$.
The contribution to $E_{0}$  is given by positive
and negative $k_{i}$ under periodic BCs, and  only by positive $k_{i}$ under zero BCs. Therefore, the values of
$E_{0}$ under zero and periodic BCs almost coincide. The difference
in these energies  is small (${\sss \triangle} E\sim E_{0}/N$) and
can be determined by the method \cite{gaudin1971} (see also \cite{batchelor2005}),
in which one should take the term $\arctan{(c/2k_{i})}$ into account.

We note that the same equations for zero BCs can be deduced by starting from (\ref{8}).

3) Strong coupling ($\gamma \gg 1$). This case corresponds to very large $c$ or $L$.
Consider Eq. (\ref{9}). The limit $c\rightarrow \infty$  means large
denominators on the right-hand side of (\ref{9}), whereas
$L\rightarrow \infty$ means small $k_{i}$, i.e., small numerators on the right-hand side of (\ref{9}).
In both cases, the sum on the right-hand side can be neglected. As a result, we obtain the solution
\begin{equation}
k_{i}=\pi i/L,
     \label{38} \end{equation}
\begin{eqnarray}
E_{0}=\sum\limits_{i=1}^{N}k_{i}^{2}= N\pi^{2}n^{2}/3.
     \label{39} \end{eqnarray}
The same solution for $E_{0}$ is obtained for the periodic
system \cite{ll1963}. This is the Girardeau limit \cite{girardeau1960}.
Formula (\ref{39}) with a subsequent correction was obtained previously
by another method  \cite{batchelor2005}.

 \section{Dispersion law}
\subsection{Zero boundary conditions}
In order to understand the meaning of an elementary excitation
for a system of
point bosons with zero BCs, let us write the system of Eqs. (\ref{8}) again:
\begin{eqnarray}
Lk_{i}&=&\pi n_{i}+\sum\limits_{j=1}^{N}\left (\arctan{\frac{c}{k_{i}-k_{j}}}\right.
\nonumber\\  &+& \left.
\arctan{\frac{c}{k_{i}+k_{j}}}\right )|_{j\neq i}, \quad i=1,\ldots, N.
     \label{8b} \end{eqnarray}
This is a system of $N$ equations, where $n_{i}$ are integers.
The ground state corresponds to $n_{i}= 1$ for all $i=1,\ldots,N$.
If at least one $n_{i}> 1$, we have an excited state.
Equations (\ref{8b}) can be compared with the keys of a piano.
The pressing of the $l$th key can be interpreted as a generation of
an elementary excitation with $n_{l}> 1$. The pressing of
the $j$th key ($j\neq l$) means the generation of the second elementary
excitation.

By minimally pressing $l$ last keys,
we obtain a configuration with  $n_{i<N+1-l}= 1$, $n_{i\geq N+1-l}= 2$, i.e.,
$l$ excitations with smallest $n_{i}>1$.
In work \cite{lieb1963}, such a structure is associated with a \textquotedblleft hole,\textquotedblright\
a second type of elementary excitations.
And excitations with $n_{i<N}= 1, n_{N}\geq  2$
are called \textquotedblleft particle states\textquotedblright \cite{lieb1963}.
We note that the analysis  \cite{lieb1963} was executed in another language, by starting from
Eqs. (\ref{4}) written for the difference $k_{i+1}-k_{i}$. However,
the properties of excitations are most  clearly seen from Eqs. (\ref{8b}).
The separation of the excitations into holes and particles is based on the analogy with a Fermi system
and can be carried out in the same way for periodic and zero BCs.
But we consider that, for a Bose system, it is more natural to describe all excitations in a unified way.
Most simply, the excitation can be associated with the clicking of a single
key. In what follows, we will define the elementary excitations namely so.

We now find the dispersion law for an elementary excitation. For the ground state,
we have  $n_{i}= 1$ for all $i$,  and some $k_{i}$ are the solutions of (\ref{8b}).
For an excited state, we write Eqs. (\ref{8b}) in the form
\begin{eqnarray}
L\acute{k}_{i}&=&\pi \acute{n}_{i}+\sum\limits_{j=1}^{N}\left (\arctan{\frac{c}{\acute{k}_{i}-\acute{k}_{j}}}
\right. \nonumber\\ &+& \left.\arctan{\frac{c}{\acute{k}_{i}+\acute{k}_{j}}}\right )|_{j\neq i}, \quad i=1,\ldots, N,
     \label{3-5} \end{eqnarray}
where $\acute{n}_{i<N}= 1$, $\acute{n}_{N}> 1$.
In this case, $ 0< \acute{k}_{1}<\acute{k}_{2}<\ldots < \acute{k}_{N}$.  We set
 \begin{equation}
\omega(k_{i})\equiv\omega_{i}=\acute{k}_{i}-k_{i}.
     \label{3-6} \end{equation}
At the transition to the excited state, only the $N$th equation in (\ref{8b}) is changed.
Therefore, we may expect \cite{lieb1963} that $\omega_{i<N}$ are small ($|\omega_{i<N}|\ll |k_{i}|$)
and $\omega_{N}$ is not small. The solution  agrees with this assumption.
In this case, the quasimomentum and the excitation energy are as follows:
 \begin{equation}
p=\sum\limits_{i=1}^{N}\omega_{i}=\int\limits_{k_{1}}^{k_{N-1}}\omega(k)\rho(k)dk  +\omega_{N},
     \label{3-7} \end{equation}
 \begin{equation}
E=\sum\limits_{i=1}^{N}(\acute{k}^{2}_{i}-k^{2}_{i})\approx \omega^{2}_{N}+2k_{N}\omega_{N}+2\int\limits_{k_{1}}^{k_{N-1}}k\omega(k)\rho(k)dk.
     \label{3-8} \end{equation}
From Eqs. (\ref{3-5}) with the numbers $i=1,\ldots, N-1,$ we now subtract
corresponding Eqs. (\ref{8b}).
In view of the smallness of $\omega_{i<N}$ and the nonsmallness of $\omega_{N}$,
we obtain the equations
\begin{eqnarray}
L\omega_{i}&=&c\sum\limits_{j=1}^{N-1}
\left \{ \frac{\omega_{j}-\omega_{i}}{(k_{i}-k_{j})^{2}+c^{2}}+\frac{-\omega_{j}-
\omega_{i}}{(k_{i}+k_{j})^{2}+c^{2}} \right \}|_{j\neq i}
\nonumber\\ &+&f(k_{i}),
     \label{3-9} \end{eqnarray}
\begin{eqnarray}
f(k_{i})&=&\arctan{\frac{c}{\acute{k}_{i}-\acute{k}_{N}}} + \arctan{\frac{c}{\acute{k}_{i}+\acute{k}_{N}}}
\nonumber \\ &-& \arctan{\frac{c}{k_{i}-k_{N}}}-\arctan{\frac{c}{k_{i}+k_{N}}},
     \label{3-10} \end{eqnarray}
$i=1,\ldots, N-1$. The similar consideration of the $N$th equations in (\ref{8b})
and (\ref{3-5}) gives the dependence
$\omega_{N}(\acute{n}_{N})$, which  is unnecessary for  finding of $E(p)$.

 In the left- and right-hand sides of (\ref{3-9}), we add the term with $i=j$, transit from summation to integration
by rule (\ref{24}), and extend the domain of definition of $\omega(k)$
to negative $k$ by the rule $\omega(-k)=-\omega(k)$. Then
relations (\ref{3-9}) yield
\begin{eqnarray}
\omega(q)\left (L-\frac{2c}{c^{2}+4q^{2}}\right )&=&
c\int\limits_{-k_{N-1}}^{k_{N-1}}dk\cdot \rho(k)
\frac{\omega(k)-\omega(q)}{(q-k)^{2}+c^{2}}
\nonumber\\ &+&f(q),
     \label{3-11} \end{eqnarray}
 $q \in [-k_{N-1},k_{N-1}]$. We  expand
(\ref{3-10}) in the small parameter $\omega_{i<N}$ and obtain
\begin{eqnarray}
f(q)&\approx &\arctan{\frac{c}{\omega_{N}+k_{N}+q}} - \arctan{\frac{c}{\omega_{N}+k_{N}-q}}
\nonumber \\ &+&
\arctan{\frac{c}{k_{N}-q}}- \arctan{\frac{c}{k_{N}+q}} \label{3-12} \\ &-&
\frac{\omega(q)c}{c^{2}+(\omega_{N}+k_{N}-q)^{2}}-\frac{\omega(q)c}{c^{2}+(\omega_{N}+k_{N}+q)^{2}}.
     \nonumber \end{eqnarray}
The sum of the first four terms in (\ref{3-12}) is larger than two
last ones by a factor of $\sim |\omega_{N}/\omega(q)| \gg 1$, and
the term $2c/(c^{2}+4q^{2})$ in (\ref{3-11}) is small as compared
with $L$ (for $\gamma \gg 1/N$). We neglect these three small
terms.

  For a periodic $(c, N, L)$-system, the relation \cite{ll1963}
\begin{equation}
2\pi\rho_{p}(k) - L=
2c\int\limits_{k_{1}=-k_{N}}^{k_{N}}dk\cdot \frac{\rho_{p}(k)}{(q-k)^{2}+c^{2}}
     \label{3-13} \end{equation}
holds, where $ 2\rho_{p}(k)=\rho(k)$ (see the previous section). Using (\ref{22}),
we obtain
\begin{eqnarray}
&&\int\limits_{-k_{N-1}}^{k_{N-1}}dk\frac{\rho(k)}{(q-k)^{2}+c^{2}} \approx
\int\limits_{-k_{N}}^{k_{N}}dk\frac{\rho(k)}{(q-k)^{2}+c^{2}} \nonumber \\ &&-
\frac{1}{c^{2}+(q-k_{N})^{2}}-\frac{1}{c^{2}+(q+k_{N})^{2}}.
     \label{3-14} \end{eqnarray}
Equalities (\ref{3-13}) and (\ref{3-14}) allow us to write (\ref{3-11}) in the form
\begin{equation}
\pi g(q)=c\int\limits_{-k_{N-1}}^{k_{N-1}}dk
\frac{g(k)}{(q-k)^{2}+c^{2}}+f(q)+\tilde{f}(q),
     \label{3-15} \end{equation}
where $g(q)=\omega(q)\rho(q),$ and
\begin{eqnarray}
\tilde{f}(q) \approx
\frac{c\omega(q)}{c^{2}+(q-k_{N})^{2}}+\frac{c\omega(q) }{c^{2}+(q+k_{N})^{2}},
     \label{3-16} \end{eqnarray}
\begin{eqnarray}
f(q)&\approx &\arctan{\frac{c}{\omega_{N}+k_{N}+q}} - \arctan{\frac{c}{\omega_{N}+k_{N}-q}}
\nonumber \\ &+& \arctan{\frac{c}{k_{N}-q}} - \arctan{\frac{c}{k_{N}+q}}.
     \label{3-17} \end{eqnarray}
In (\ref{3-15}), the function $\tilde{f}(q)$ can be neglected, because $|\tilde{f}(q)|\ll |f(q)|$.
Since $f(-q)=-f(q)$, relation (\ref{3-15}) yields $g(-q)=-g(q)$.
We note that the same equations are obtained, if we determine
$\acute{k}_{i}-k_{i}$ from Eqs. (\ref{9}) and (\ref{10}).

\begin{figure}
\includegraphics[width=0.47\textwidth]{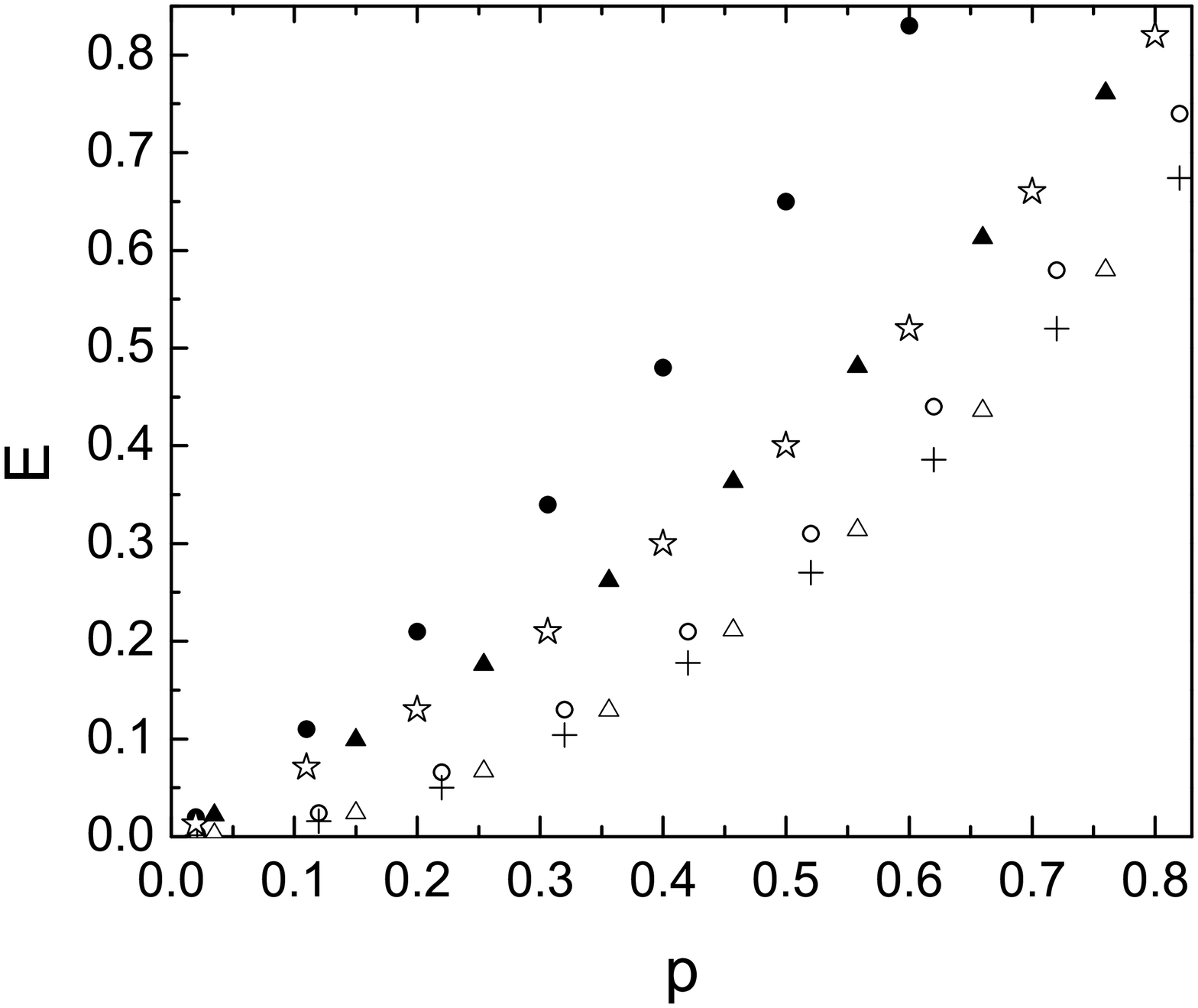}
\hfill
\includegraphics[width=0.47\textwidth]{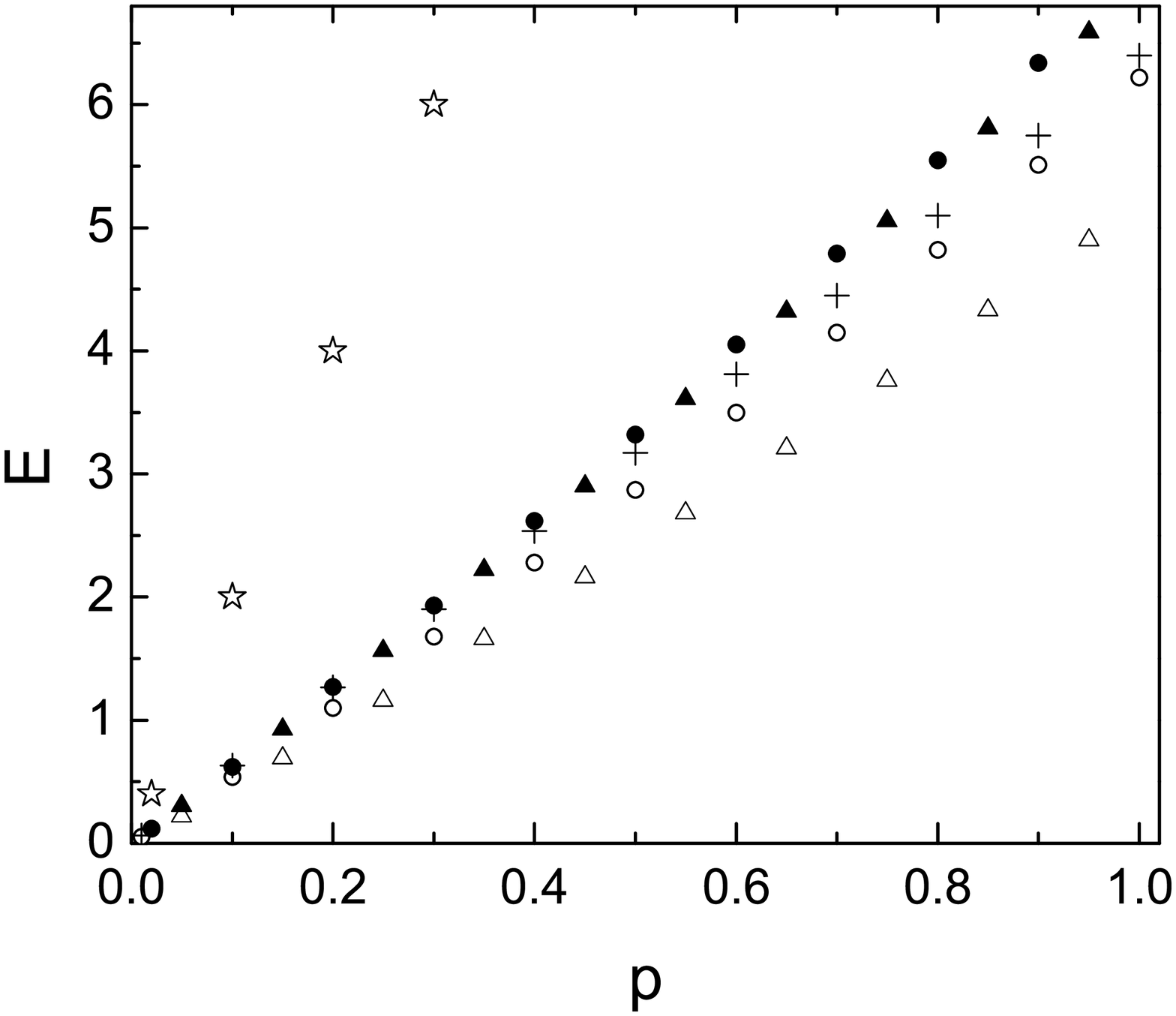}
\\
\parbox[t]{0.47\textwidth}{
\caption{ Dispersion laws for point bosons for $n=1$ and 1) $\gamma=c=0.001$:
the curves in the cases of zero (open circles) and periodic
(open triangles) BCs and the Bogolyubov law (crosses);
2) $\gamma=c=0.1$: zero BCs (circles), periodic BCs (triangles), and the Bogolyubov law (stars).
 \label{fig1}} } \hfill
\parbox[t]{0.47\textwidth}{
\caption{ Dispersion laws for  $n=1$ and 1) $\gamma=c=10$:
the curves in the cases of zero (open circles) and periodic
(open triangles) BCs and the Bogolyubov law (crosses);
2) $\gamma=c=100$: zero BCs (circles), periodic BCs (triangles),
and the Bogolyubov law (stars).  \label{fig2}} }
\end{figure}

The dispersion law $E(p)$ can be obtained by formulae (\ref{3-7})
and (\ref{3-8}) with $k_{N-1}\approx k_{N}-1/\rho(k_{N})$, if we consider $\omega_{N}$ as a
free parameter varying from $0$ to $\infty$ and, for each $\omega_{N},$ find the function $g(q)$
from Eqs. (\ref{3-15}) and (\ref{3-17}). The quantities  $k_{N}$ and $\rho(k_{N})$ follow from the
Lieb--Liniger equations for the ground state \cite{ll1963} (Eq. (\ref{23}) with
$\rho(k)\rightarrow \rho_{p}(k)$, $k_{1}=-k_{N}$, and Eq. (\ref{3-13})).
Equations (\ref{3-15}) and (\ref{3-17}) can be easily solved numerically.
For the replacement of the integral in (\ref{3-15}) by a sum,  we must merely make sure that
the step is sufficiently small,  so that at small $c$ the inequality $|q-k^{n}_{j}|< c$
will be valid for at least ten numerical points $k^{n}_{j}$.

The solutions for $E(p)$ are presented in Figs. 1--3 in comparison with the solutions for a periodic system
\cite{lieb1963} and the Bogolyubov law \cite{bog1947,bz1956} for the point potential
\begin{equation}
E_{b}(p)= \sqrt{p^{4}+4cnp^{2}}.
     \label{3-18} \end{equation}
By comparing Fig. 3 with Figs. 1 and 2, we see the dependence  of
$E(p)$ on $n$ for the same $\gamma$.
In the limit $p\rightarrow \infty,$ all curves approach the
asymptotics $E=p^{2}$. It is seen from the figures that, for
$\gamma = c/n\ll 1,$  the curve for periodic BCs is close to the
Bogolyubov law, but the curve for zero BCs is noticeably
\textit{different} from it. In particular, the effective sound
velocity $(E/p)|_{p \rightarrow 0}$ for zero BCs is larger than
that by Bogolyubov by $1.56$ times for $\gamma=0.001$ and $ n=1;
100$ and by $1.49$ times for $\gamma=0.1$ and $n=1; 100$ (our numerical calculation for $\gamma =0.01$, $n=1$
 gave for periodic BCs the sound velocity to be $0.97$ of
the Bogolyubov one). As $\gamma$ increases, the curves for periodic and
zero BCs approach each other. As $\gamma \gg 1,$ they
are close to the Girardeau curve
$E_{g}(p)=p^{2}+2\pi np$ \cite{girardeau1960}.
The relation $\gamma \gg 1$ means $L\gg 1$ or $c\gg 1.$  It is easily seen from
Eqs. (\ref{9}) and (\ref{10}) that, in these cases, $k_{i}\approx
\pi I_{i}^{z}/L\approx \acute{k}_{i}$ for $i<N$, which yields
$\omega(k) \rightarrow 0$. Therefore, we obtain
from (\ref{3-7}) and (\ref{3-8}) $p=\omega_{N}, E(p)=p^{2}+2k_{N}p$. Since
$k_{N}|_{\gamma \rightarrow \infty}\rightarrow \pi n$
\cite{ll1963}, we have the Girardeau law. So, this law holds
at $c=\infty$ \cite{girardeau1960} and  $c\lsim 1, \gamma \gg 1$ \cite{lieb1963}  for
 periodic BCs and  at  $ \gamma \gg 1$ for zero BCs.

The solution $E(p)$ for zero BCs can be written in the Bogolyubov form (\ref{3-18}) with the replacement
$c\rightarrow c\cdot\vartheta(k,c,n)$, where $\vartheta(k,c,n)$ depends weakly on $k$ and strongly on $c$ and $n$.

We note that, for $\gamma \lsim 1$ and small or intermediate $E$ and $p,$
the main contribution to $E$ and $p$ is given by small perturbations $\omega_{i<N}$. For large $E$ and $p,$
the main contribution to these quantities
is given by $\omega_{N}$. In other words, the excitations are collective
for small  $E$ and $p,$  and are quasi-one-particle for  large $E$ and $p$.
For $\gamma \gg 1,$ the excitations are quasi-one-particle for any $E$ and $p$, even for $p,\omega_{N}\rightarrow 0$.
For all curves, $p_{\omega_{N}\rightarrow 0}\rightarrow 0$ for any $c$ and $n$.
In Fig. 4, we show the dependence $p(\omega_{N})$ (\ref{3-7}) under zero and periodic BCs.

\begin{figure}
\includegraphics[width=0.47\textwidth]{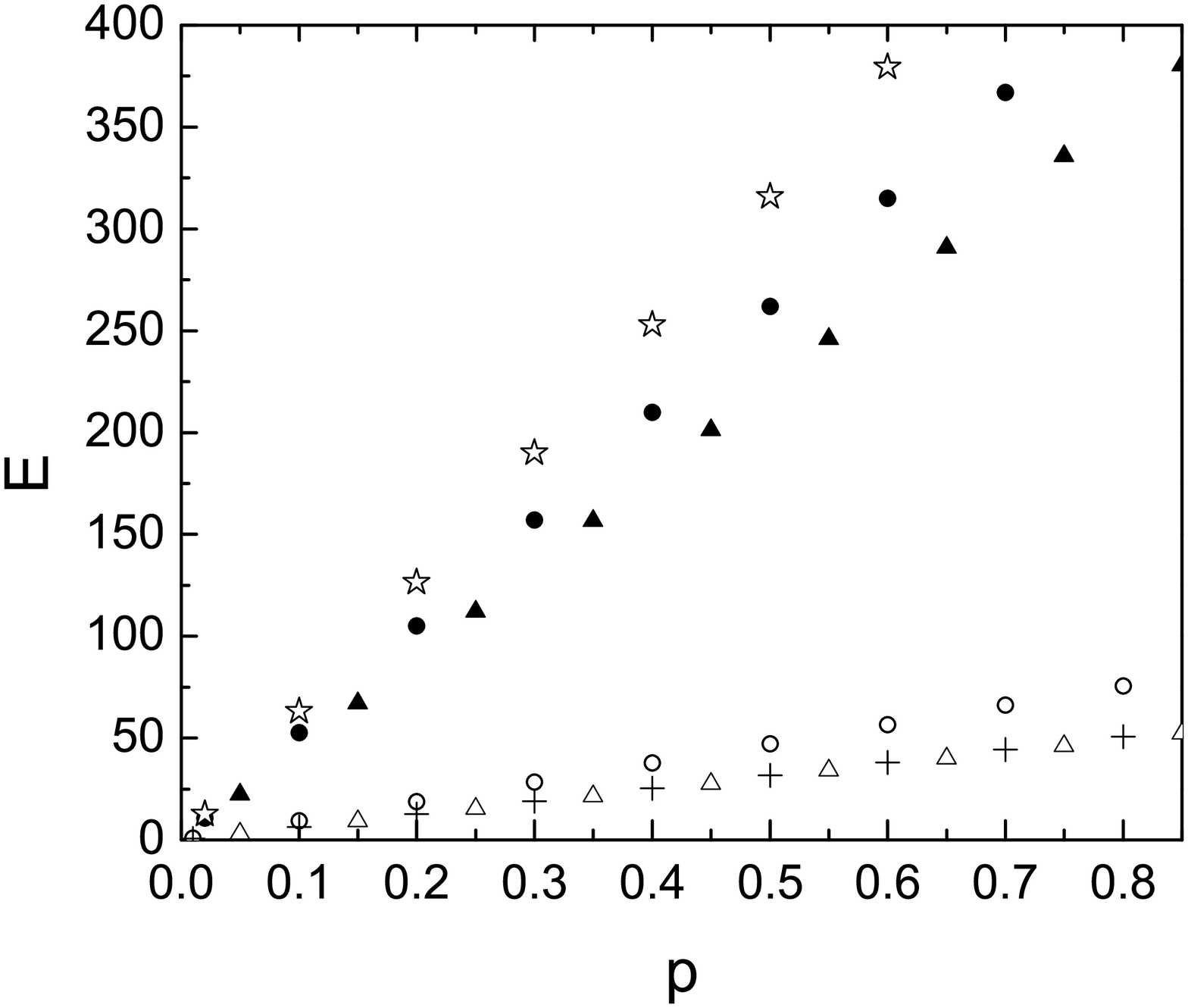}
\hfill
\includegraphics[width=0.47\textwidth]{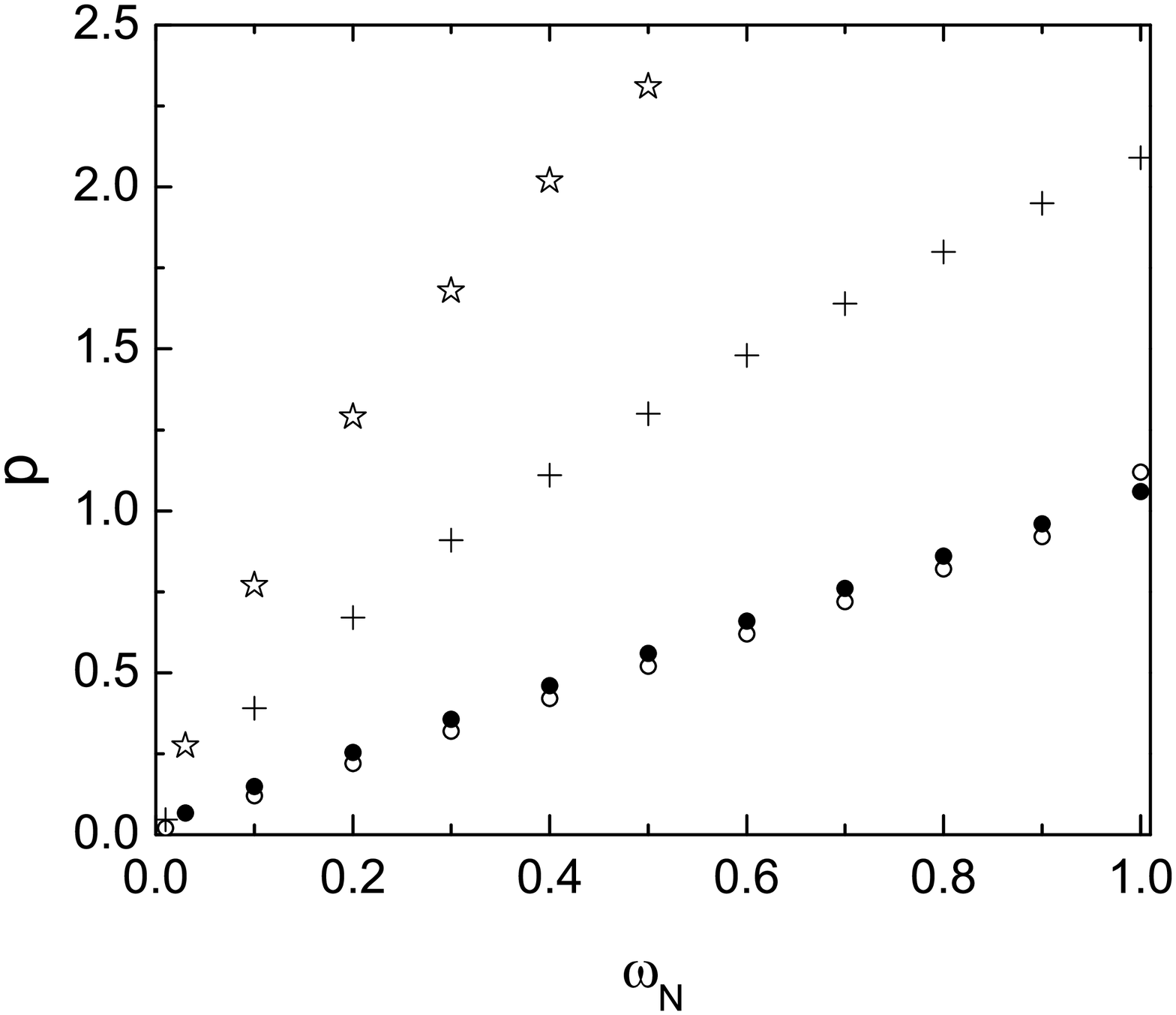}
\\
\parbox[t]{0.47\textwidth}{
\caption{ Dispersion laws for  $n=100$. 1) $c=10$, $\gamma=0.1$:
the solutions in the cases of zero (open circles) and periodic  (open triangles) BCs and the Bogolyubov law (crosses).
2) $c=1000$, $\gamma=10$:  zero BCs (circles), periodic BCs  (triangles), and the Bogolyubov law (stars).
 \label{fig1}} } \hfill
\parbox[t]{0.47\textwidth}{
\caption{  Dependence $p(\omega_{N})$ (\ref{3-7}) for $\gamma=0.001$:
1) zero BCs and $n=1$, $c=0.001$ (open circles);
2) zero BCs and $n=100$, $c=0.1$ (crosses);
3)  periodic BCs and $n=1$, $c=0.001$ (circles);
4)  periodic BCs and $n=100$, $c=0.1$ (stars).
 \label{fig4}} }
\end{figure}

\subsection{Periodic boundary conditions}
The dispersion law for periodic BCs was found by E.H.~Lieb \cite{lieb1963}.
In order to understand the reason for the influence of boundaries on the dispersion law, let us compare the
formulae obtained for zero and periodic BCs. For a periodic $(c, N, L)$-system,
the equations are deduced exactly in the same way as in the case of zero BCs.
Starting from Eqs.
(\ref{4}) and (\ref{5}) or from the equivalent equation (\ref{7}), we obtain
\begin{equation}
2\pi g_{p}(q)=2c\int\limits_{k_{1}=-k_{N}}^{k_{N}}dk
\frac{g_{p}(k)}{(q-k)^{2}+c^{2}}+f_{p}(q),
     \label{3-20} \end{equation}
\begin{eqnarray}
f_{p}(q)&\approx &2\arctan{\frac{\omega_{N}+k_{N}-q}{c}}\nonumber\\&-&2\arctan{\frac{k_{N}-q}{c}},
     \label{3-21} \end{eqnarray}
where $g_{p}(q)=\omega(q)\rho_{p}(q)$.
Equations (\ref{3-7}) and (\ref{3-8}) remain valid if we
replace $\rho(q) \rightarrow \rho_{p}(q)$ and consider $k_{N-1}\approx k_{N}-1/\rho_{p}(k_{N})$.
The values of $\rho_{p}(q)$ and $k_{N}$ can be obtained from the Lieb--Liniger equations \cite{ll1963}
 for the ground state of a periodic system (Eq. (\ref{23}) with
$\rho(k)\rightarrow \rho_{p}(k)$, $k_{1}=-k_{N}$, and Eq. (\ref{3-13})).
In Figs. 1-3, we give the dispersion laws  for periodic BCs, which were obtained numerically
from Eqs. (\ref{3-7}), (\ref{3-8}), (\ref{3-20}), and (\ref{3-21})
with the indicated changes.

In the derivation of Eqs. (\ref{3-20}) and (\ref{3-21}), we considered $k_{1},\ldots,k_{N-1}$
to be the quasimomenta  of a system of $N$
interacting atoms (like under zero BCs).
 Eqs. (2.18)--(2.20) from \cite{lieb1963} were obtained within another approach,
where  $k_{1},\ldots,k_{N-1}$ were considered to be the quasimomenta of a system of $N-1$
interacting atoms. In the first (second) approach, at $\omega_{N}>0$
we have  $\omega(q)> 0$ ($\omega(q)< 0$).
The advantage of the second approach is that the derivation of equations is simpler.
However, the ground state corresponds  to $N$
interacting atoms. Therefore, the first approach is slightly more exact.
But the results in both approaches are very close.
The first approach has the advantage that the properties $E(\omega_{N}\rightarrow 0)\rightarrow 0,
p(\omega_{N}\rightarrow 0)\rightarrow 0$ follow
directly from the input equations  (\ref{3-7}), (\ref{3-8}), (\ref{3-20}), and (\ref{3-21}),
whereas the analogous properties in the second approach are not obvious and require a
bulky proof \cite{lieb1963}.

\section{Numerical solution by the Newton method}
Trying to solve equations of the form (\ref{8}) numerically, we found that
this can be performed  easily (for $N\lsim 1000$) and with a high accuracy   within the Newton method.
This method frequently requires the proximity of a bare  solution to the exact one. However, for systems
(\ref{7}) and (\ref{8}), the method converges also with the quick choice of a bare  solution.
The essence of the method is as follows. Two nonlinear equations
\begin{equation}
f_{1}(k_{1},k_{2})=0, \quad f_{2}(k_{1},k_{2})=0
     \label{50-1} \end{equation}
can be approximately written in the form
\begin{eqnarray}
&&f_{1}(k^{(0)}_{1},k^{(0)}_{2})+\frac{\partial f_{1}}{\partial k_{1}}(k^{(0)}_{1},k^{(0)}_{2})\cdot (k_{1}-k^{(0)}_{1})
\nonumber \\ &+&
\frac{\partial f_{1}}{\partial k_{2}}(k^{(0)}_{1},k^{(0)}_{2})\cdot (k_{2}-k^{(0)}_{2})=0,
     \label{50-2} \end{eqnarray}
analogously for  $f_{2}$.  Setting $k^{(0)}_{j}=k^{(l-1)}_{j}$, $k_{j}=k^{(l)}_{j}$ ($j=1, 2$),
we get a linear recurrence relation between $k^{(l)}_{j}$ and $k^{(l-1)}_{j}$. Under certain conditions,
the collection $\{k^{(l)}_{j}\}$ converges with increasing $l$ to the exact solution $\{k_{j}\}$.
\begin{figure}[ht]
\centerline{\includegraphics[width=85mm]{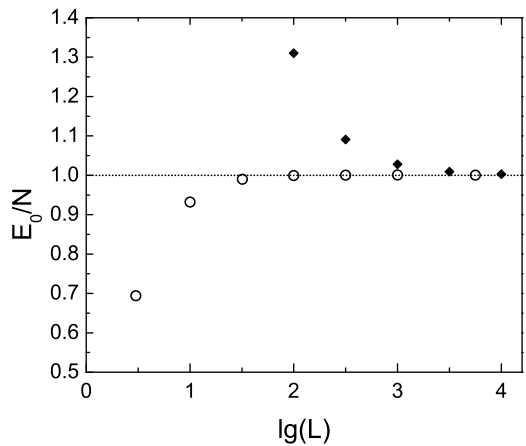}
}
\caption{ Dependence of $E_{0}/N$ on the system size $L$ under zero (rhombs)
and periodic (circles) BCs for $n=1$, $\gamma=0.01$. The value of $E_{0}$
is obtained by direct solving of the systems of equations (\ref{7}) and (\ref{8}) within the Newton method.
The dotted line shows the Bogolyubov
solution $E_{0}/N=cn(1-4\sqrt{\gamma}/3\pi)$ taken as 1 ($E_{0}/N=1$).
 \label{fig5}}
\end{figure}

By this method, we found $E_{0}$ and $E(k)$
for periodic and zero BCs by means of the direct solving of Eqs.
(\ref{7}) and (\ref{8}). To within $1\% ,$ the solutions
coincide with the above-obtained ones.
We get also the dependence $E_{0}(L)$ for $n=const$
(see Fig. 5). The difference in the values of $E_{0}$ for small $L$ from $E_{0}(L\rightarrow \infty)$
is the finite-size effect. The similar effect was previously studied  in \cite{ying2001,batchelor2005}.

The Newton method is a powerful tool for solving of systems of the form (\ref{4}), (\ref{7}),  (\ref{8}), (\ref{9}).

\section{Thermodynamics}
In the following section, we will see that  the excitations are not phonons.
Therefore, we cannot observe them by the scattering of some particles.
It is of interest to clarify whether the difference in dispersion curves for zero and periodic BCs
leads to a difference in the measured thermodynamic quantities.
C.N. Yang and C.P. Yang  \cite{yangs1969} constructed the
thermodynamic description of point bosons, by using the mixed language of atoms
and excitations (see also \cite{jiang2015}).  In particular,
the free energy has the Fermi form \cite{yangs1969}
\begin{equation}
F = N\mu-\frac{k_{B}TL}{2\pi \hbar}\int\limits_{-\infty}^{\infty}dp
\ln{\left (1+e^{\frac{-\epsilon(p)}{k_{B}T}}\right )},
     \label{6-0} \end{equation}
where $\epsilon(p)$ is some effective energy
that is not the energy of a quasiparticle.  To study the influence of the boundaries, it is necessary to
carry on the whole analysis \cite{yangs1969} anew for zero BCs.
However, the weakly excited state of a quantum liquid can be considered as  a number of excitations and
is most simply described  in the language of excitations \cite{land9}.
Moreover, the excitations are usually observed rather than atoms.
If we separate the excitations into ``holes''
and ``particles'', then the excitations do not obey
some simple statistics. The ``mixed'' description was constructed
\cite{yangs1969} probably just for this reason.
Below, we propose a simpler way to construct the thermodynamics,
which will allow us to introduce excitations in a self-consistent manner.

In what follows, the formulae are valid for zero and periodic BCs.
The Gibbs canonical distribution implies that the free energy of the system reads \cite{land5}
\begin{equation}
F= -k_{B}T\ln{\sum\limits_{j}e^{-E_{j}/k_{B}T}},
     \label{6-1} \end{equation}
where $j$ enumerates all possible states of the system, and $E_{j}$ is the energy of the system in the $j$th
state.  Any excited state of the system is uniquely determined by the set of numbers
$\{n_{i}\}\equiv (n_{1},n_{2},\ldots,n_{N})$ in (\ref{7}) or (\ref{8}).
For the ground state, we have the set $(n^{(0)},\ldots,n^{(0)})$. Let in system (\ref{7}) or (\ref{8})
the number of equations with $n_{l}\neq n^{(0)}$ be much less than $N$. The analysis by the
Newton method shows that, in this case, the energy of the system is
\begin{equation}
E(\{n_{i}\})\approx E_{0}+\sum\limits_{i=1}^{N}\varepsilon_{i}(l_{i}),
     \label{6-2} \end{equation}
 \begin{equation}
\varepsilon_{i}(l_{i})=E(n^{(0)},\ldots,n^{(0)},n_{i},n^{(0)},\ldots,n^{(0)})-E_{0},
     \label{6-3} \end{equation}
where $l_{i}=n_{i}-n^{(0)}$, and $i$ is the number of the equation in system (\ref{7}) or (\ref{8}).
Because all  $i$ are equivalent, we have $\varepsilon_{i}(l_{i})=\varepsilon(l_{i})$.
The quantity $\varepsilon(l_{i})$
coincides with the \textit{energy of a quasiparticle } (\ref{3-8}). $l_{i}$ can take the zero value, then
$\varepsilon(0)=0$.
Then relation (\ref{6-1}) can be written in the form
\begin{equation}
F= E_{0} -k_{B}T\ln{\sum\limits_{l_{1}\ldots l_{N}}e^{-\sum\limits_{i=1}^{N}\varepsilon(l_{i})/k_{B}T}}.
     \label{6-4} \end{equation}
In sum (\ref{6-4}), we should take only different states into account \cite{land5}.
For example, all states of the form $n_{i\neq j}=n^{(0)}, n_{j}=2$
are equivalent and must be considered as a one state.
It is difficult to determine such a sum (\ref{6-4}) if $N$ is finite.
However, this can be easily performed in the limit
$N=\infty, L=\infty, N/L=n$. For  infinite $N,$ the different states from the
set $(l_{1},\ldots ,l_{N})$ can be enumerated by the set of numbers $\{ \eta_{l}\}$,
where $\eta_{l}$  is the occupation number for the $l$th state,
and $l$ has the same values as any $l_{i}$. It is also necessary to replace
\begin{equation}
\sum\limits_{i=1}^{N}\varepsilon(l_{i})|_{N\rightarrow\infty}\rightarrow \sum\limits_{l}\eta_{l}\varepsilon(l),
     \label{6-5} \end{equation}
where the numbers $\eta_{l}$ for each $l$ can take the values $\eta_{l}=0,1,2,\ldots, \infty$.
  As a result, we have
\begin{equation}
F= E_{0} -k_{B}T\ln{\sum\limits_{ \{\eta_{l}\}}e^{-\sum\limits_{l}\eta_{l}\varepsilon(l)/k_{B}T}},
     \label{6-6} \end{equation}
where $\{\eta_{l}\}$ is the set $\eta_{1}, \eta_{2},\ldots, \eta_{l},\ldots$ (the enumeration coincides with
that one for $l$, see below). Let us rewrite $F$ in the form
\begin{equation}
F= E_{0} -k_{B}T\sum\limits_{l}\ln{\sum\limits_{ \eta_{l}=0}^{\infty}e^{-\eta_{l}\varepsilon(l)/k_{B}T}}.
     \label{6-7} \end{equation}
 Summing the geometric progression, we obtain finally
\begin{equation}
F= E_{0} +k_{B}T\sum\limits_{l}\ln{\left (1-e^{-\varepsilon(l)/k_{B}T}\right )}.
     \label{6-8} \end{equation}
This formula describes the free energy of a system of noninteracting bosons \cite{land5}
with zero chemical potential and the additional summand $E_{0}$.
Formula (\ref{6-8}) indicates that the weakly excited
state of a system of point bosons can be considered as a number of
elementary excitations satisfying the \textit{Bose} statistics.

This corresponds to the symmetry of wave functions.  Indeed, the permutation of the $l$th and  $j$th
excitations means the permutation of $n_{l}$
and  $n_{j}$ in Eq. (\ref{7}) or (\ref{8}). This leads only to the permutation of
$k_{i}$ in the complete collection $\{ k_{i}\}$. In this case, the
Lieb--Liniger \cite{ll1963} and Gaudin \cite{gaudin1971} wave functions are invariable.
In addition, the system can possess several excitations with identical $n_{l}$.
These properties again indicates that
the excitations are \textit{bosons} for any $\gamma$ and $k$.
By their properties, the excitations are similar to Bogolyubov ones \cite{bog1947,bz1956}.

We note that equality (\ref{6-2}) holds if the total number of excitations $\ll N$, i.e., if $T$ is low.
If the number of excitations is of the order of magnitude of $ N$, one needs to consider their interaction.

For zero BCs, the levels are numbered as follows:  $l$ can take the values $1, 2, \ldots$
 corresponding to the quasimomenta
$k^{(l)}=(\sum\limits_{i}k_{i})|_{(n_{j\leq N-1}=1,n_{N}=l+1)}-(\sum\limits_{i}k_{i})|_{(n_{j\leq N}=1)}$.
For periodic BCs, we have $l=\pm 1, \pm 2,  \ldots$,  $k^{(l)}=(\sum\limits_{i}k_{i})|_{(n_{j\leq N-1}=0,n_{N}=l)}-
(\sum\limits_{i}k_{i})|_{(n_{j\leq N}=0)}$,  $k^{(-l)}=-k^{(l)}$.
By the Newton method, we determined the first $1000$ levels $(\varepsilon(l),k^{(l)})$
for periodic and zero BCs for $N=1000, n=1,$
and $\gamma =0.01; 1$.  It turns out that $\varepsilon_{z}(2l)=\varepsilon_{p}(l)=\varepsilon_{p}(-l)$,\
$\varepsilon_{z}(2l-1)=0.5(\varepsilon_{p}(l)+\varepsilon_{p}(l-1))$
(the indices $p$ and $z$ mean periodic and zero BCs,
and we consider $\varepsilon_{p}(l=0)=0$).
These equalities hold with accuracy $\sim 1/N$. If we neglect the small difference $\varepsilon_{p}(l)-\varepsilon_{p}(l-1)$,
we obtain $\varepsilon_{z}(1)=\varepsilon_{p}(-1),\ \varepsilon_{z}(2)=\varepsilon_{p}(1),\
\varepsilon_{z}(3)=\varepsilon_{p}(-2),\ \varepsilon_{z}(4)=\varepsilon_{p}(2),$ and so on.
In other words, the energies of levels under periodic and zero BCs \textit{coincide}. Therefore, the free energies
(\ref{6-4}) for the systems under periodic and zero BCs are \textit{identical}.
With regard to the difference $\varepsilon_{p}(l)-\varepsilon_{p}(l-1)$,
we obtain a surface correction $\sim F/N$ to the free energy $F$.

In (\ref{6-8}), we may pass to the integration with respect to $p$ or $E$:
\begin{equation}
F \approx E_{0} +\frac{k_{B}TL}{2\pi \hbar}\int\limits_{-\infty}^{\infty}dp \zeta(p) \ln{\left (1-e^{\frac{-E(|p|)}{k_{B}T}}\right )},
     \label{6-9} \end{equation}
\begin{equation}
F \approx E_{0} +\frac{k_{B}TL}{2\pi \hbar}\int\limits_{-\infty}^{\infty}dE \tilde{\zeta}(E) \ln{\left (1-e^{\frac{-E}{k_{B}T}}\right )},
     \label{6-10} \end{equation}
where $\zeta(p_{l})=2\pi[L(k^{(l+1)}-k^{(l)})]^{-1}$ for periodic BCs and $\zeta(p_{l})=\pi [L(k^{(l+1)}-k^{(l)})]^{-1}$
for zero BCs, $p_{l}=\hbar k^{(l)}$,  $\tilde{\zeta}(E)=\zeta(p)\cdot\partial p/\partial E$.
For zero BCs, $\zeta(-|p|)\equiv \zeta(|p|)$.
The numerical analysis indicates that, for periodic BCs,
 $k^{(l+1)}-k^{(l)}=2\pi/L$ and $\zeta(p)=1$. For zero BCs, the step $k^{(l+1)}-k^{(l)}$ depends on $l$
 and differs from the step for periodic BCs (which leads to the different dispersion law). In this case,
$\tilde{\zeta}(E)$ is the same under periodic and zero BCs with deviation $\lsim 0.3\%$.
Therefore, relation (\ref{6-10}) allows us again to conclude that $F$ is independent of the boundaries.

Formula (\ref{6-9}) with $\zeta(p)=1$ describes the free energy
of a gas of excitations in He II \cite{xal} (with the additional term $E_{0}$ independent of $T$).

It is of interest that the total number of excitations in a gas of point particles is
at most $N,$ by definition. For a gas of nonpoint particles, the excitation is manifested as a multiplier of the total
wave function \cite{bz1956,fc,yuv1980,zero-liquid}, and the number of multipliers is unbounded.
Moreover, the system of energy levels of $N$ point particles \textit{has no}
level corresponding to $N+1$ phonons in a gas of $N$ nonpoint particles.
That is, some  states of a gas of nonpoint particles have an analog in
a gas of point particles, whereas another states do not have.
Intuitively it seems  that a real system of $100$ Bose particles
can hold $150$ oscillatory waves (phonons). It is possible that,
for finite $N,$ the excitations for point particles are defined not quite self-consistently.

\section{Nature of excitations, open questions}
We have shown above that, for a weak and intermediate couplings, 
the dispersion law for point bosons does depend on the presence
of boundaries, whereas the ground-state energy is independent of boundaries
(to within the surface correction $\sim E_{0}/N$).
Why is it so as regards equations? As was shown in Sec. III,
the equations for the ground state for systems with zero and periodic BCs differ from each
other only by two terms, which are small for large $N$ and $L$.
But the equations for the dispersion law differ from each other strongly.
Namely, for periodic BCs, the functions  $f_{p}(q)$ and $g_{p}(q)$ in Eqs. (\ref{3-20}), (\ref{3-21})
are constant-sign for all $q$.
But, under zero BCs, $f(-q)=-f(q)$ and $g(-q)=-g(q)$ (see Eqs. (\ref{3-15}), (\ref{3-17})).
Thus, the equations under periodic and zero BCs  \textit{differ by their symmetries}.
Therefore, the solutions for the  dispersion laws are also different.

However, for zero BCs, the low-lying excitations are not phonons.
In order to see this, let us compare
the microscopic sound velocity
$v_{s}^{micro}=\partial E(p)/\partial p |_{p\rightarrow 0}$
with the macroscopic one  $v_{s}^{macro}=\sqrt{\partial P/\partial \rho},
P=-\partial E_{0}/ \partial V$ $(\rho = mn)$ \cite{putterman}.
Under the periodic BCs, they are identical \cite{girardeau1960,lieb1963}.
Therefore, the excitations with small $p$ can be interpreted as phonons.
Under zero BCs, the system is characterized by the same $E_{0}$ and $v_{s}^{macro}$,
but by the different  value of $v_{s}^{micro}$, being approximately  $1.5$ times over $v_{s}^{macro}$ at $\gamma \ll 1$.
Hence, under zero BCs, the excitations with small $p$  are not  phonons.
This is true for $\gamma \lsim 10$.
For $100 \lsim \gamma < \infty,$ the relation $v_{s}^{micro}\approx v_{s}^{macro}$ holds,
and the excitations are almost phonons.
For $\gamma=\infty,$ we have $v_{s}^{micro}= v_{s}^{macro}$, and the excitations can be considered as phonons.
Which structure of the wave function (WF) under zero BCs should be in order that an excitation be a phonon?
The total WFs are not eigenfunctions of the operator of total momentum even for $\gamma=\infty$.
But the WF can contain a multiplier corresponding to two counter-propagating waves.
In this case, the WF of a low-energy state should have the form
\begin{equation}
 \Psi(x_{1},\ldots, x_{N})=(\psi(x_{1},\ldots, x_{N}|k)+\psi(x_{1},\ldots, x_{N}|-k))\Psi_{0},
      \label{5-1} \end{equation}
where the function $\psi(x_{1},\ldots, x_{N}|k)$ is an eigenfunction of the total operator of momentum
with the eigenvalue $k$. The structure of (\ref{5-1}) is phonon-like.
If such representation is possible, then the
excited state has the phonon structure and is characterized by the quasimomentum $k$.
In this case, the relation $v_{s}^{micro}= v_{s}^{macro}$ should hold.
But since $v_{s}^{macro}$ is the same as for periodic BCs, the dispersion law $E(k)$
should coincide with that for a periodic system. We do not know whether a representation of the form
(\ref{5-1}) exists.

According to solutions \cite{zero-liquid,zero-gas}, the boundaries of a system of nonpoint bosons
affect both the dispersion law and the ground-state energy.
For a 1D system  of almost point bosons with the weak interaction and zero BCs,
the following dispersion law is found \cite{zero-liquid,zero-gas}:
\begin{equation}
E(p)= \sqrt{p^{4}+2cnp^{2}}.
     \label{5-18} \end{equation}
For the solutions \cite{zero-liquid},   the equality
$v_{s}^{micro}= v_{s}^{macro}$ holds; this can be verified for the weak coupling.
The dispersion law (\ref{5-18}) is characterized by the sound velocity, which is
$\sqrt{2}$ times less than the Bogolyubov one (see (\ref{3-18})).
However, for the point bosons in a box, the effective $v_{s}^{micro}$ is larger than the
Bogolyubov $v_{s}$. If there exists a continuous transition from a nonpoint interaction
to the point one, then the noncoincidence of solutions \cite{zero-liquid}  with those for point
bosons indicates the incorrectness of either solutions \cite{zero-liquid} or the solutions of the present work.
However, we solved the Gaudin equations for $k_{i}$ by two different methods
and are sure in the validity of the solutions.
In addition, if there is no phonon representation for excitations of a gas of point
bosons under zero BCs, we are faced with the difficulty for the theory of point bosons.
Indeed, for the real quantum Bose liquids, the low-lying excitations
are phonons. This is testified, for example, by the experiments on the scattering of
neutrons and by the measurements of the heat capacity of 3D He II in a vessel (zero BCs).
The distinction of the one- and three-dimensional cases should not be important,
because several microscopic models of He II \cite{bz1956,fc,yuv1980,zero-liquid}
work in 1D and 3D and give for 1D and 3D the solutions of the same structure.

As a possible reason for all disagreements, we can indicate
the absence of a continuous transition from
a nonpoint interaction to the point one, i.e.,
the anomality of the $\delta$-function.
The $\delta$-function is a singular generalized function.
It is commonly accepted that the replacement of a real potential by the $\delta$-function
is admissible. In particular, it was proved mathematically  \cite{seiringer2008}
that the energy levels of a system of 3D bosons in a very extended trap are close to those of the Lieb--Liniger problem.
However, the wave functions of $N$ nonpoint and $N$ point particles have different forms \cite{zero-liquid}.
Is it the different forms of the same functions or the evidence of the difference of the functions?
It is known only \cite{girardeau1960} that, under periodic BCs, the WFs of  point bosons with $\gamma=\infty$
can be written as the zero approximation for the WFs of nonpoint bosons.
It is necessary to show that, for an $N$-particle 1D system, all  energy levels and the WFs for the almost
point and point interactions coincide. In the Appendix, this is proved for the one-particle problem.
The same  should be proved at least for $N=2$ as well.
It is of interest  that, in the 2D- and 3D-spaces,
the potential $\delta(\textbf{x})$ has no influence on the solutions
of the Schr\"{o}dinger equation  for some tasks \cite{simenog2014}.

The equations for spin systems and  point bosons are similar  \cite{gaudinm}.
Therefore, one can expect that the dispersion laws of
spin waves under zero and periodic BCs should be different
 (by our method of determination of $k$),
whereas the thermodynamic quantities should coincide. From whence,
we may conclude that the difference of the curves $E(k)$ is  unobservable.
But it was shown in experiments on the scattering of neutrons that
the low-lying excitations of magnetics with boundaries (zero BCs)  are quite observable and have
quasimomentum. The possible reason is that the spin wave is accompanied by the sound wave.
An important point is that the contact   Hamiltonian describes well the real exchange interaction
and contains no $\delta$-function.
Therefore, the solutions should correspond to the natural properties,
and  disagreements due to the $\delta$-function should not arise.
Such a deductive method indicates that the problems of solutions for spin waves and point bosons
under zero BCs are not, apparently, related to the $\delta$-function.

 \section{Conclusion}
We have obtained two main results:  1)
It is found that the dispersion law $E(k)$ of a system of point bosons depends strongly on boundaries
in the regimes of weak and intermediate coupling.
2) The thermodynamics of a gas of point bosons is constructed by a new method.
Our analysis shows that the values of thermodynamic quantities are independent of the boundaries.
By our method of determination of the quasimomentum $k$ of an excitation, it turns out that, under zero BCs,
the low-energy excitations are characterized by a linear dispersion law and a
nonphonon structure of the wave function. It seems strange, because the experiment indicates
that the low-lying excitations of real uniform quantum liquids with zero boundary conditions are phonons.
It is possible that there exists a way of determination of $k$ under zero BCs,
for which the low-energy excitations are phonons.
Otherwise, the solutions for point bosons do not describe the real
low-lying modes, and we meet an internal difficulty of theory.

The author is grateful to Yu. V. Shtanov for the discussion and the indication
of an error in the analysis of the $\delta$-function. 
I  also thank the  referees for  helpful remarks.

 \section{Appendix. Comparison of the solutions for point and almost point potentials}
The interatomic potentials $U(r)$ are usually have
a high repulsive barrier in the region $r\lsim 2\,\mbox{\AA}$ and a shallow pit in the region
$2\,\mbox{\AA} \lsim r \lsim 5\,\mbox{\AA}$ and tend
asymptotically to zero, as $r$ increases \cite{szalewicz2012}. Is it possible to model
a nonpoint high barrier with the $\delta$-function? To answer,
 we compare the solutions for the wave functions and the
energies obtained for both potentials.

Here, we consider the simplest one-particle task: a particle in the one-dimensional
potential well $-L/2\leq x \leq L/2$ (i.e., zero BCs at $x=\pm L/2$)
 with the potential barrier
 \begin{equation}
 U(|x|) =
\left [ \begin{array}{ccc}
    \bar{U}=R_{0}U_{0}/a> 0,  & \   |x|\leq a,   & \\
    0,  & \ |x|>a &
\label{4-1} \end{array} \right. \end{equation}
at the well center. Here, $a\ll L/2$. The Fourier transform of such potential is $\nu(k)=(2R_{0}U_{0}\sin{ak})/ak$.
By passing to the limit $a\rightarrow 0,$
we have $\nu(k)\rightarrow 2R_{0}U_{0}=2c=const$, which corresponds to  $U(x)=2c \delta(x)$.
Let us compare the solutions for almost point  (arbitrarily small, but finite $a$) and point interactions.

The Schr\"{o}dinger equation reads
\begin{equation}
 -\Psi''(x)+U(x)\Psi(x)=E\Psi(x).
     \label{4-2} \end{equation}
At a finite $a,$ we seek a solution in the form
\begin{equation}
 \Psi(x) =
\left [ \begin{array}{ccc}
    b_{1}\cos{kx}+b_{2}\sin{kx},  &    x\in[-L/2,-a[,   & \\
    d_{1}e^{-\kappa x}+d_{2}e^{\kappa x},  & \   |x|\leq a,   &  \\
    a_{1}\cos{kx}+a_{2}\sin{kx},  & \   x\in ]a,L/2]. &
 \label{4-3} \end{array} \right. \end{equation} Relation
(\ref{4-2}) yields $E=k^{2}, \kappa=\sqrt{\bar{U}-k^{2}}$  (for
$|k|\leq \sqrt{\bar{U}}$). The boundary conditions and the sewing
conditions for $\Psi(x)$ and $\Psi'(x)$ yield the equations
\begin{equation}
 b_{1}\cos{(kL/2)}-b_{2}\sin{(kL/2)}=0,
     \label{4-4} \end{equation}
\begin{equation}
 a_{1}\cos{(kL/2)}+a_{2}\sin{(kL/2)}=0,
     \label{4-5} \end{equation}
\begin{eqnarray}
 b_{1}\cos{ka}-b_{2}\sin{ka}=d_{1}e^{\kappa a}+d_{2}e^{-\kappa a},
     \label{4-6} \end{eqnarray}
\begin{eqnarray}
 b_{1}k\sin{ka}+b_{2}k\cos{ka}=-\kappa d_{1}e^{\kappa a}+\kappa d_{2}e^{-\kappa a},
     \label{4-7} \end{eqnarray}
\begin{eqnarray}
 a_{1}\cos{ka}+a_{2}\sin{ka}=d_{1}e^{-\kappa a}+d_{2}e^{\kappa a},
     \label{4-8} \end{eqnarray}
\begin{eqnarray}
 -a_{1}k\sin{ka}+a_{2}k\cos{ka}=-\kappa d_{1}e^{-\kappa a}+\kappa d_{2}e^{\kappa a}.
     \label{4-9} \end{eqnarray}
They have two \textquotedblleft branches\textquotedblright\ of solutions. For $a\rightarrow 0$ and
 $k^{2}\ll \bar{U},$  we have \\
I) $d_{1}=d_{2}=a_{1}/2$, $b_{1}=a_{1}$, and $b_{2}=-a_{2}=a_{1}/\tan{(kL/2)}.$
The value of $a_{1}$ can be found from the normalization condition, and $k$ satisfies the equation
 \begin{equation}
 \tan{(kL/2)}=-k/c.
     \label{4-10} \end{equation}
II) $d_{1}=-d_{2}=-ka_{2}/2\kappa \rightarrow 0$, $b_{1}=-a_{1}=0$, $b_{2}=a_{2}$, $a_{2}$ is determined
from the normalization condition, and $k$ satisfies the equation
 \begin{equation}
 \tan{(kL/2)}=0, \ k= 2\pi l/L, \ l= 1,  2,  3, \ldots
     \label{4-11} \end{equation}

For the point potential $U(x)=2c \delta(x),$ we seek the solution of Eq. (\ref{4-2})
in the form (\ref{4-3}) without the second row. We possess the BCs
\[\Psi(\pm L/2)=0,\]
the condition of continuity of $\Psi(x)$ at the point of the barrier
\[\Psi(x=-\delta)|_{\delta \rightarrow 0}=\Psi(x=\delta)|_{\delta \rightarrow 0},\]
and the equation
\begin{equation}
 [\Psi'(x=\delta) - \Psi'(x=-\delta)]|_{\delta \rightarrow 0}=2c\Psi(0),
\label{4-13} \end{equation}
 obtained by the integration of
the Schr\"{o}dinger equation (\ref{4-2}) on the interval $x\in
[-\delta, \delta]$ (a similar equation arises also for $N$ point
bosons \cite{ll1963}). These equations have two branches of
the solutions: 1) $E=k^{2}$, $b_{1}=a_{1}$,
$b_{2}=-a_{2}=a_{1}/\tan{(kL/2)}$, and Eq. (\ref{4-10}) for $k$; 2)
$E=k^{2}$, $b_{1}=a_{1}=0$, $b_{2}=a_{2}$, and Eq. (\ref{4-11}) for $k$.
They  coincide with solutions (I) and (II) for almost point particles.
We note that though the function $\delta(x)$ does not act on odd functions,
such functions can be eigenfunctions of the Hamiltonian with the  $\delta$-function.

Let us consider the properties of solutions.  For series (I),
the lower level corresponds to the WF without nodes.
The next levels correspond to the WFs with two, four, etc nodes.
For series (II), the lower level corresponds to the WF with a
single node. For the next levels, the WFs have three, five, etc
nodes. By the theorem of nodes \cite{gilbert}, the ground state
corresponds to the WF without nodes, the first excited state
to the WF with one node, the second  excited state to the WF
with two nodes, etc. Solutions (I) and (II) correspond to the theorem of nodes.

Note that the eigenvalue $E=k^{2}$ of the Schr\"{o}dinger equation coincides with the value of $\langle
\Psi_{0}|\hat{H}| \Psi_{0}\rangle$ both for the almost point
interaction and  for the point one.
In the proof, it is necessary to consider that,
for the point potential, $\Psi_{0}'(x)$ has a discontinuity at the point $x=0$  (see (\ref{4-13})).

     \renewcommand\refname{}

       \end{document}